\documentclass[runningheads]{llncs}
\usepackage{graphicx}
\usepackage{float}
\usepackage{hyperref}
\usepackage{courier}
\usepackage{listings}
\usepackage{tikz}
\def\checkmark{\tikz\fill[scale=0.4](0,.35) -- (.25,0) -- (1,.7) -- (.25,.15) -- cycle;} 

\usepackage{multirow}

\begin{document}

\title{Using Mapping Languages for Building Legal Knowledge
Graphs from XML files\thanks{Copyright $\copyright$ 2019 for this paper by its authors. Use permitted under Creative Commons License Attribution 4.0 International (CC BY 4.0).}}
%
%
\author{Ademar Crotti Junior\inst{1}\and
Fabrizio Orlandi\inst{1}\and
Declan O'Sullivan\inst{1}\and 
Christian Dirschl\inst{2}
\and Quentin Reul\inst{3}}
\authorrunning{Crotti Junior, A. et al.}
\titlerunning{Using Mapping Languages for Building Legal Knowledge
Graphs}
%
\institute{ADAPT Centre for Digital Content Platform Research, Knowledge \& Data Engineering Group, School of Computer Science and Statistics, Trinity College Dublin, Dublin 2, Ireland \\
\email{\{ademar.crotti,fabrizio.orlandi,declan.osullivan\}@adaptcentre.ie} \and
Wolters Kluwer Deutschland, München, Germany \\
\email{christian.dirschl@wolterskluwer.com} \and
Wolters Kluwer N.V., Chicago, USA
\\
\email{quentin.reul@wolterskluwer.com}
}
\maketitle              
\begin{abstract}
This paper presents our experience on building RDF knowledge graphs for an industrial use case in the legal domain. The information contained in legal information systems are often accessed through simple keyword interfaces and presented as a simple list of hits. In order to improve search accuracy one may avail of knowledge graphs, where the semantics of the data can be made explicit. Significant research effort has been invested in the area of building knowledge graphs from semi-structured text documents, such as XML, with the prevailing approach being the use of mapping languages. In this paper, we present a semantic model for representing legal documents together with an industrial use case. We also present a set of use case requirements based on the proposed semantic model, which are used to compare and discuss the use of state-of-the-art mapping languages for building knowledge graphs for legal data.
\end{abstract}

\section{Introduction}\label{sec:introduction}

The body of law to which citizens and businesses have to adhere is constantly increasing in volume and complexity \cite{Boella2016}. The information contained in such a body of law is usually provided by  unstructured text within legal documents, for which a number of systems have been developed. The information made available by such legal information systems, however, is often accessed with simple, keyword-based search interfaces and presented as a simple list of hits \cite{Filtz2017}. This makes the process of information retrieval time consuming and inefficient, especially when dealing with large amounts of information \cite{Schweighofer2010}. Moreover, the usefulness of such information varies widely and depends on its structure and its representation. In this context, although the information may be available, users and legal professionals may find the exploration of legal information problematic when interested in specific circumstances or investigating a particular case \cite{Schweighofer2010}. Such issues have led to a need for improving ways to search and structure large amounts of legal information. 

This work presents ongoing efforts related to building RDF knowledge graphs for representing legal documents. The key focus of building such knowledge graphs is to improve search accuracy by understanding its intent and context. The RDF (Resource Description Framework) \cite{cyganiak2014rdf} data model is used here as it is a W3C Recommendation which allows one to describe resources and their relationships by the means of vocabularies and ontologies in a way that computerized agents are able to process. Ontologies, in this context, are seen as formal, explicit specifications of conceptualizations \cite{gruber1995toward}. The structure and semantics provided by ontologies allows one to formulate complex questions such as \textit{“"What are the documents in which a relation to a particular law concept, or a more specific one, exists?"”}.


In this paper, we present a semantic model for legal documents, which is then being used in a real-world use case. This use case comes from an ongoing project with Wolters Kluwer Germany, where legal documents must be transformed to RDF knowledge graphs. These documents are stored as XML files and follow a specific schema. Considering the semantic model and the legal document's schema we have defined a set of requirements, which are used to compare and discuss the use of different state-of-the-art mappings engines. Finally, we present an evaluation comparing the performance of a mapping approach and an \textit{ad hoc} custom parser.

The remainder of this paper is organised as follows: Section \ref{sec:use-case} describes our use case and the XML document schema. Section \ref{sec:semantic-model} presents a semantic model for representing legal documents that has been developed. Section \ref{sec:semantic-uplift} describes the semantic uplift of legal documents to RDF knowledge graphs through the use of mapping languages. Section \ref{sec:evaluation} presents a comparison evaluation between mapping engines. Section \ref{sec:relatedwork} discusses related work. Section \ref{sec:conc} concludes the paper and discusses future work.

\section{Use case}\label{sec:use-case}

This section presents our industrial use case, which comes from an ongoing project with Wolters Kluwer Germany (WKD). WKD is a leading knowledge and information service provider in the domains of law, companies, and tax, which offers high quality business information for professionals. Wolters Kluwer is based in more than 40 countries and serves customers in more than 180 countries worldwide.

\begin{figure}[H]
\caption{Excerpt from a WKD's XML legal document}
\includegraphics[width=12cm]{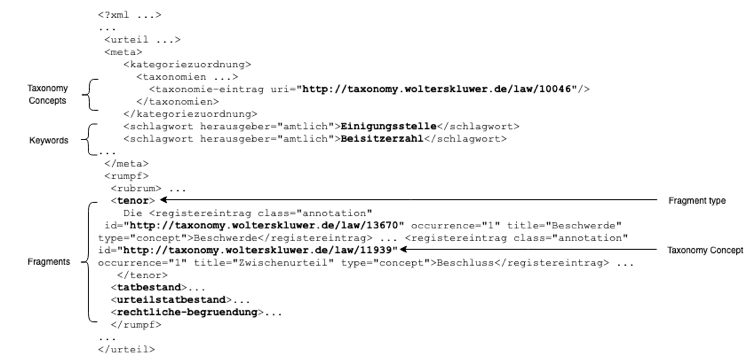}
\centering
\label{fig:doc}
\end{figure}

WKD's use case contains millions of documents in the German language containing legal information together with links to taxonomy concepts. As mentioned, the documents are stored as XML files. Each document consists of the key parts: document keywords, document taxonomy concepts, and fragments. Each fragment has a type, such as main claim (\textit{tenor}), court facts (\textit{tatbestand}), amongst others, which are represented by different XML element tags. Just like documents, fragments may be annotated with taxonomy concepts. The use of taxonomy concepts defines the specific legal matters and processes contained in a document. In this sense, the shared use of such concepts across documents reflect the relations between the legal information contained in those documents. This characteristic, however, is not made explicit, since each document is represented by a single different XML file. One possible way of making such information explicit is through the use of knowledge graphs, as will be discussed in Section 3. An example of an XML document is shown in Figure \ref{fig:doc}.

The taxonomy concepts used to annotate documents and its fragments come from WKD's taxonomy ontology. This ontology contains information about legal concepts, which are supplemented by technical terms of neighboring areas such as economics, sociology or politics. The Simple Knowledge Organization System \footnote{\url{https://www.w3.org/TR/skos-reference/}} (SKOS) vocabulary is used to describe concepts and their relations in this taxonomy. SKOS is a W3C Recommendation designed to support the use of knowledge organization systems. In WKD's use case, each legal concept is represented as a \texttt{skos:Concept}, with the main relationships being expressed through the properties \texttt{skos:narrower} and \texttt{skos:broader}. A major legal subdomain of WKD's taxonomy is available for download and accessible via a SPARQL endpoint\footnote{\url{http://taxonomy.wolterskluwer.de/}}.

\section{A Semantic Model for Legal Documents}\label{sec:semantic-model}

As stated in Section \ref{sec:use-case}, our use case shares taxonomy concepts within and across legal documents, which are not made explicit through the XML data format. This section presents a semantic model designed for the representation of those legal documents, with the aim of making such relationships explicit.

\begin{figure}[H]
\caption{Semantic model for legal documents}
\includegraphics[width=13cm]{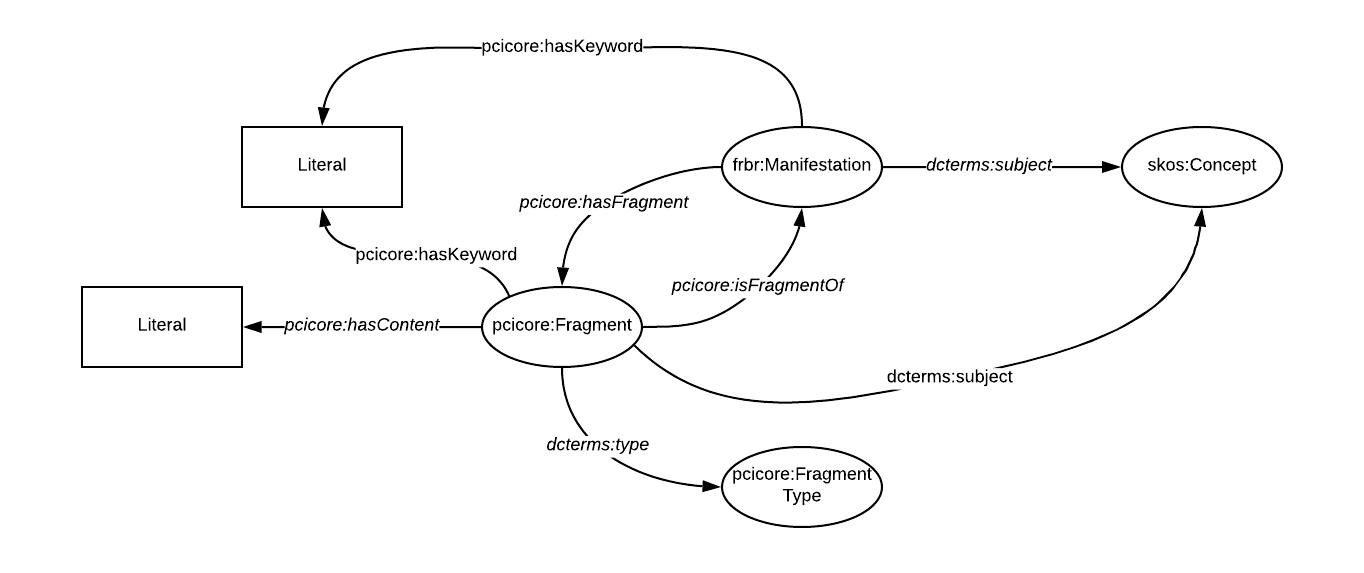}
\centering
\label{fig:semantic-model}
\end{figure}

The proposed semantic model draws inspiration and extends an existing one presented in \cite{6142232}. Our semantic model is also leveraged by WKD's taxonomy concepts, which are used to link entire documents and fragments to the legal concepts defined in the taxonomy. Figure \ref{fig:semantic-model} shows the proposed semantic model used to represent the legal information contained in our use case (Section \ref{sec:use-case}). 

The vocabularies being used in the semantic model are the Platform Content Interface (PCI), which leverages the Functional Requirements for Bibliographic Records (FRBR), the Simple Knowledge Organization System (SKOS) and the Dublin Core (DC) terms ontologies. The PCI ontology is a proprietary vocabulary describing legal documents and metadata. The FRBR\footnote{\url{http://purl.org/vocab/frbr/core}} ontology provides a vocabulary for concepts and relations in bibliographic databases defined by the International Federation of Library Associations\footnote{\url{https://www.ifla.org/}} initiative. Finally, the DC terms\footnote{\url{http://purl.org/dc/terms/}}  ontology provides a vocabulary describing all metadata terms maintained by the Dublin Core Metadata Initiative\footnote{\url{http://dublincore.org/}}. The PCI ontology was also extended in order to represent the content of fragments within a document with the datatype property \texttt{pcicore:hasContent}, and with the object property \texttt{pcicore:isFragmentOf}, which is also defined as an inverse property of \texttt{pcicore:hasFragment}. The latter two properties allow one to reference documents to its fragments, and vice-versa. 

Each document is represented as a \texttt{frbr:Manifestation}. The keywords of the document are described with the datatype property \texttt{pcicore:hasKeyword}, and the related taxonomy concepts with the property \texttt{dcterms:subject}. Each fragment is represented as a \texttt{pcicore:Fragment}, and includes a reference to the document it belongs to through the property \texttt{pcicore:isFragmentOf}. The content of fragments are represented with \texttt{pcicore:hasContent}. Fragments have a type represented with the class \texttt{pcicore:FragmentType}. Each fragment may also have keywords (\texttt{pcicore:hasKeyword}), and be annotated with taxonomy concepts (\texttt{dcterms:subject}). The legal taxonomy concepts, which are described as instances of the class \texttt{skos:Concept}, provide linkable anchors both to entire documents and to smaller fragments. In this context, concepts are used to connect legal documents and textual pieces of supporting evidence within and across different documents.

As discussed in Section \ref{sec:introduction}, the representation of legal information through knowledge graphs allows one to formulate complex questions. An example, which was stated in Section 1, is the question: \textit{“"What are the documents in which a relation to a particular law concept, or a more specific one, exists?"”}. This question can be answered with the SPARQL query presented in Listing \ref{list:sparql}. Note that this query returns documents related to a specific concept (in this case \texttt{wkd-law:10046}) at either the document or fragment level, and that the property \texttt{skos:narrower} is used to refer to more specific concepts from the taxonomy ontology.

\begin{lstlisting}[language=sparql, caption=Example of SPARQL query, frame = single, label={list:sparql}]
PREFIX pcicore: <http://onto.wolterskluwer.com/pci/core/>
PREFIX skos: <http://www.w3.org/2004/02/skos/core#>
PREFIX wkd-law: <http://taxonomy.wolterskluwer.de/law/> 
PREFIX dcterms: <http://purl.org/dc/terms/>

SELECT distinct ?document
WHERE { BIND (wkd-law:10046 as ?concept) 
    ?fragment a pcicore:Fragment; pcicore:isFragmentOf ?document.
    { ?fragment dcterms:subject ?concept . } 
    UNION { ?fragment dcterms:subject ?narrower .  ?concept skos:narrower ?narrower . } 
    UNION { ?document dcterms:subject ?concept .  } 
    UNION { ?document dcterms:subject ?narrower . ?concept skos:narrower ?narrower . }
}
\end{lstlisting}



\section{Semantic Uplift}\label{sec:semantic-uplift}

This section presents our use case requirements, which are defined based on the XML document’s schema and the proposed semantic model, together with a comparison between state-of-the-art semantic uplift engines applied to our use case.

Several approaches have been developed in the area of semantic uplift through mapping languages. Mapping languages can be described as declarative languages used to express customized mappings defining how non-RDF data should be represented in RDF \cite{DBLP:journals/ijwis/JuniorDBO17}. An engine is usually associated with a mapping language, being a software processor that uses a mapping file and the input data to generate RDF datasets. A mapping file contains one or more mapping definitions, which state how the RDF terms are generated, considering the input data, the vocabularies being used, and how these are associated to each other.

\subsection{Use Case Requirements}

In our case, the following requirements must be met by the semantic uplift engine in order to transform the XML files (Section 2) into the semantic data model (Section 3)\footnote{These requirements express both generic requirements for building knowledge graphs as well as specific ones for our use case, such as being capable to select and transform XML attributes to RDF resources.}.

\begin{itemize}

\item \textbf{R1. Data format.} This requirement is related to legal documents in our use case being stored as XML files, being also a common data format used in many applications.

\item \textbf{R2. Data selection.} This requirement is related to selecting specific XML elements and attributes during the mapping process. We note that this includes the mapping of elements which contain other nested XML elements.

\item \textbf{R3. Vocabulary independent.} This requirement allows the mapping to be defined using existing ontologies and vocabularies.

\item \textbf{R4. Transformation functions.} This requirement allows for values to be manipulated during the mapping process. For instance, in our use case, some elements in the XML documents contain string values that must be normalized in order to be represented in RDF.

\item \textbf{R5. Multi-attribute mapping.} This requirement is related to the XML files in our use case having a collection of value nodes that are mapped to one property in the RDF representation.

\item \textbf{R6. Literal values to IRI.} This requirement is related to IRIs being stored as literals in the XML documents, which are required to be transformed into valid IRIs in the RDF representation.

\end{itemize}

\subsection{Semantic Uplift Engines}

The following semantic uplift engines were compared when considering our use case. The rationale for selecting these being that, according to their specification, they would have support for our use case requirements. 

\textbf{XSPARQL.} XSPARQL \cite{Bischof2012} is a query language combining XQuery and SPARQL for transformations between RDF and XML (lifting) and back (lowering). For the former, XSPARQL uses a combination of XQuery expressions and SPARQL CONSTRUCT queries. The XQuery expressions are used to access XML data, and the SPARQL CONSTRUCT queries are used to convert the accessed XML data to RDF. For the later, XSPARQL uses a combination of SPARQL and XQuery clauses. The SPARQL clauses are used to access RDF data, and the XQuery clauses are used to format the results in XML syntax. This combination of languages allows one to benefit from the facilities of SPARQL for retrieving RDF data, and the use of a TURTLE like syntax for constructing RDF graphs, while still having access to XQuery features for XML processing. Transformation functions in XSPARQL are supported through native functions found in SPARQL, XQuery, XPath and XSLT.

\textbf{SPARQL-Generate.} SPARQL-Generate \cite{DBLP:conf/esws/LefrancoisZB17} extends SPARQL with specific target constructs which enable the generation of RDF from heterogeneous sources. SPARQL-Generate supports the generation of RDF from any RDF dataset, and from any set of documents in arbitrary formats, such as XML, CSV and so on. SPARQL-Generate has been designed as an extension of SPARQL 1.1, which means that it can be implemented on top of any existing SPARQL engine by leveraging the SPARQL extension mechanism to deal with an open set of formats. In order to do so, SPARQL-Generate introduced three clauses to their SPARQL extension. The source clause is used to reference the input source data. The iterator clause allows for the extraction of data attributes from a given source data. These attributes are then bound to SPARQL variables. Finally, the generate clause replaces and extends the SPARQL CONSTRUCT clause with SPARQL-Generate queries. The bounded variables which refer to data attributes are used here to form the RDF triples. SPARQL-Generate supports data transformation functions through native SPARQL 1.1 functions.

\textbf{RML-Mapper.} R2RML\footnote{\url{https://www.w3.org/TR/r2rml/}} is the W3C standardized mapping language for defining mappings of data in relational databases to the RDF data model. The RML \cite{dimou_ldow_2014} extension of R2RML broadens its scope by also covering the (semi-) structured formats CSV, XML and JSON. RML documents contain rules defining how the input data will be represented in RDF. The main building blocks of R2RML and RML mapping documents are Triples Maps. A Triples Map defines how the RDF triples of the form (subject, predicate, object) will be generated. A Triples Map consists of one Logical Source, one Subject Map and zero or more Predicate-Object Maps. The Subject Map defines how identifiers (IRIs) are generated for the mapped resources, which are used as the subject of the RDF triples. A Predicate-Object Map consists of Predicate Maps, which define how to generate the triple’s predicate and Object Maps or Referencing Object Maps, which define how the triple's object is generated. The Subject Map, the Predicate Map and the Object Map may be called Term Maps. Term Maps express how an RDF term – which may be an IRI, a blank node or a literal – is generated. A Term Map can be a constant-valued term map which is always generating the same RDF term, a reference-valued term map that is the data value of a referenced attribute from a given Logical Source, or a template-valued term map that is a valid string template that may contain referenced attributes from a given Logical Source. The engine RML-Mapper supports data transformation functions through the Function Ontology \cite{DBLP:conf/esws/MeesterDVM16}. 

\textbf{CARML.} CARML\footnote{\url{https://github.com/carml/carml}} is an engine which implements the RML mapping language, just like the described RML-Mapper. In this sense, CARML also supports the generation of RDF datasets from heterogeneous data formats. Transformation functions are also supported through the Function Ontology.

\subsection{Discussion}

In order to compare these engines, a mapping expressing the transformations needed to convert the XML files, as described in Section 2, to the RDF semantic model described in Section \ref{sec:semantic-model} was created for each of these mapping engines. In order words, this comparison assesses if and how well such engines support the our use case requirements.

Table \ref{table-engines} shows the mapping engines being evaluated in our use case, their licenses, and the version used. A discussion on the support for each use case requirement is presented next.

\begin{table}[H]
\centering
\begin{tabular}{|l|l|l|}
\hline
\textbf{Mapping Engine} & \textbf{License} & \textbf{Version} \\ \hline
XSPARQL & BDS                   & 4.0.0 \\ \hline
SPARQL-Generate  & Apache &    1.3.1 \\ \hline
RML-Mapper & MIT & 4.3.3 \\ \hline
CARML & MIT & 0.2.3 \\ \hline
\end{tabular}
\label{table-engines}
\caption{\label{table-engines}Mapping engines}
\end{table}






\textbf{R1. Data format.} All the mapping engines presented have support for the conversion of XML files to RDF. XSPARQL uses XQuery, XPath and XSLT in order to access the information contained in XML files. The SPARQL-Generate, RML-Mapper and CARML engines rely on XPath expressions in order to access the data contained in XML files.

\textbf{R2. Data selection.} All of the mapping engines have support for selecting XML attributes. In order to select XML elements in XSPARQL one may use the XPath function \texttt{text()} or \texttt{string()}. The function \texttt{text()} returns a set of individual nodes contained in an XML element. For instance, if an XML element contains one nested XML element, then this function returns two nodes. The function \texttt{string()} returns the string value, or the string representation, of an XML element. In other words, an element with a nested element would return one string value containing the whole string within that XML element. The mapping engines SPARQL-Generate and RML-Mapper only allow for the selection of XML elements using the \texttt{text()} function, which means that instead of one string representation of an XML element the engine produces a set of literals for an ontology property. In our use case, this is problematic when mapping the content of fragments –- represented in the XML files by an element, often containing nested elements -- to the property \texttt{pcicore:hasContent}. CARML, on the other hand, allows for the selection of elements using both \texttt{text()} and \texttt{string()} functions. Thus, only XSPARQL and CARML fully support this requirement.

\textbf{R3. Vocabulary independent.} All the engines have support for this requirement, being expressive enough for the definition of customized mappings. XSPARQL and SPARQL-Generate have a similar syntax based on SPARQL CONSTRUCT queries to define how the RDF triples are generated from XML files. The RML-Mapper and CARML engines rely on Triples Maps, which as stated previously, allows one to define how subjects, predicates and objects are generated from non-RDF source data.

\textbf{R4. Transformation functions.} XSPARQL partially supports data transformation functions, being limited to the expressiveness of SPARQL, XQuery, XPath and XSLT. SPARQL-Generate also partially supports data transformation functions, being limited to the ones supported in SPARQL 1.1. The RML-Mapper and CARML approaches, as previously stated, fully support data transformation functions through the Function Ontology. In our use case, data transformation functions are required when mapping string values to literals where such values must be normalized and validated.

\textbf{R5. Multi-attribute mapping.} All of the engines support this requirement. The XSPARQL engine, however, has the word \textit{uri} as part of its grammar. This word is also the name of an attribute in the XML files in our use case, which results in an error in the execution of the mapping.

\textbf{R6. String values to IRI.} All of the engines support this requirement. XSPARQL allows one to define an IRI from a string value by enclosing the variable representing the IRI with less ($<$) and greater ($>$) than symbols. SPARQL-Generate allows the same transformation through the SPARQL function URI. The RML-Mapper and CARML approaches allows such transformation by defining an IRI term type to Term Maps.

Table \ref{table-reqs} presents the support for each requirement by the mapping engines. CARML is the only engine with full support for all use case requirements (Section 2)\footnote{A \checkmark means full support, while a (\checkmark) means partial support.}.

\begin{table}[H]
\centering
\begin{tabular}{|l|l|l|l|l|l|l|}
\hline
\textbf{Mapping Engine} & \textbf{R1} & \textbf{R2} & \textbf{R3} & \textbf{R4} & \textbf{R5} & \textbf{R6} \\ \hline
XSPARQL                 & \checkmark             &   \checkmark          &   \checkmark          &    (\checkmark)         &   (\checkmark)          &    \checkmark         \\ \hline
SPARQL-Generate         &     \checkmark        &    (\checkmark)         &   \checkmark          &    (\checkmark)         &     \checkmark        &     \checkmark        \\ \hline
RML-Mapper              &     \checkmark        &   (\checkmark)          &    \checkmark         &     \checkmark        &     \checkmark        &     \checkmark        \\ \hline
CARML                   &      \checkmark       &    \checkmark         &    \checkmark         &     \checkmark        &      \checkmark       &     \checkmark        \\ \hline
\end{tabular}
\label{table-reqs}
\caption{\label{table-reqs} Mapping engines comparison.}
\end{table}

\section{Evaluation}\label{sec:evaluation}

This section presents an evaluation comparing the performance of two approaches applied to the semantic uplift of XML files. One approach utilizes RML mappings through the CARML engine, which, as discussed, is the only one with full support for our use case requirements. The second approach is an \textit{ad hoc} custom parser developed to generate the same RDF representation from XML files. 

The experiment was executed on a MacBook Pro 13" (3.1 GHz, i7, 16GB RAM), where both approaches would transform the same input into the same RDF representation. Each approach was executed 10 times considering 3 different   datasets containing 1000 (1k), 10000 (10k) and 50000 (50k) documents. These datasets have been created by randomly selecting files from our use case which contains over 1 million documents. These were selected randomly in order to provide sets of documents with different characteristics (e.g. size). Table \ref{table-eval} presents the results of this experiment.

\begin{table}[H]
\centering
\begin{tabular}{|l|c|l|c|l|c|l|}
\hline
\multirow{2}{*}{\textbf{\begin{tabular}[c]{@{}l@{}}Performance\\ Comparison\end{tabular}}} & \multicolumn{2}{c|}{\textbf{1k}}                              & \multicolumn{2}{c|}{\textbf{10k}}                             & \multicolumn{2}{c|}{\textbf{50k}}                              \\ \cline{2-7} 
                                                                                           & \textbf{AVG}              & \multicolumn{1}{c|}{\textbf{STD}} & \textbf{AVG}              & \multicolumn{1}{c|}{\textbf{STD}} & \textbf{AVG}               & \multicolumn{1}{c|}{\textbf{STD}} \\ \hline
\textbf{\textit{Ad hoc} custom parser}                                                                            & \multicolumn{1}{l|}{4.09} & 0.47                              & \multicolumn{1}{l|}{38.7} & 3.95                              & \multicolumn{1}{l|}{212.8} & 30.01                             \\ \hline
\textbf{CARML}                                                                           & \multicolumn{1}{l|}{4.85} & 1.33                              & \multicolumn{1}{l|}{43.3} & 3.9                               & \multicolumn{1}{l|}{242.5} & 30.9                              \\ \hline
\end{tabular}
\label{table-eval}
\caption{\label{table-eval} Time performance results for 10 runs (in seconds).}
\end{table}


These results show that the use of mappings, namely RML with the CARML engine, does impact performance. In order to assess whether these differences are statistically significant we performed the Welch Two Sample T-Test with a significance level of 0.05. The results suggest no statistical significance between using an \textit{ad hoc} custom parser and CARML for 1k documents (p-value of 0.12). However, for 10k and 50k documents the difference in performance is statistically significant (p-values of 0.04 and 0.01, respectively). We do note that, even though the use of mapping languages impact performance, this is still the preferred approach for our use case. The reason being that mapping languages separate mapping definitions from the implementations that execute them, allowing the process to be reused and shared. Furthermore, the maintenance of mappings is facilitated by the same reason when compared to an \textit{ad hoc} parser. For instance, any changes in the semantic model or in the input data would require the mapping to be updated accordingly, without the need to change the engine responsible for the execution of the mapping. An \textit{ad hoc} parser, on the other hand, would require the implementation, which is specific to a certain use case, to be modified. Finally, implementations may be improved with optimizations and other choices of software libraries, which could be tailored to different use cases and thus positively impact performance.

\section{Related work}\label{sec:relatedwork}
The representation of legal information through ontologies have been the focus of several studies.

Ebenhoch \cite{ebenhoch2001legal} has proposed the representation of legal information through RDF, where a key point to make such data more accessible is the process of enriching it with metadata. Winkels et al. \cite{Winkels} describe the need of semantics in a legal context from the practical point of view of the Dutch Tax and Customs Administration, who have to deal with legal information from various sources and formats. JURION \cite{10.1007/978-3-319-34129-3_40} was a legal information platform which merges and interlinks over one million documents of content and data from diverse sources, such as national and European legislation and court judgments, amongst others. The approaches presented in  \cite{ebenhoch2001legal} and \cite{Winkels} used \textit{ad hoc} parsers for the creation of knowledge graphs from source data, while JURION's extraction process relies on XSLT scripts in order to convert the data stored as XML files to RDF. This paper, in contrast, has investigated the use of mapping languages for building knowledge graphs to represent legal documents.

Other studies have focused on investigating ontology design patterns in the legal domain which are described together with examples \cite{griffo2016pattern}. A summary of existing legal ontologies is provided in \cite{Breuker:2009:FCD:1563987.1563990}. This work describes and classifies 23 legal ontologies in distinct categories such in which type of application the ontology has been used, how the ontology was constructed, its language, and so on. Semantic Web technologies are also being used to describe particular subdomains of law, such as licenses, and more recently the General Data Protection Regulation \cite{DBLP:conf/esws/PanditFOL18}. As stated previously, a semantic model for legal documents has been presented in \cite{6142232}, which was used as the base for our semantic model (Section \ref{sec:use-case}).

\section{Conclusions and future work}\label{sec:conc}

In this paper, we have presented a semantic model for legal documents applied to an industrial use case. This use case comes from Wolters Kluwer Germany's company, where their legal information is stored as XML files. By taking into account the proposed semantic model and the XML's document schema we have defined a set of requirements which should be met by the mapping engine in order for it to be able to produce the required RDF knowledge graph. We have also compared and discussed the use of four different state-of-the-art mapping engines based on our set of requirements. CARML, which implements the RML mapping language, was found to be the only one supporting all our use case requirements. Finally, we have also compared and discuss the use of CARML to developing \textit{ad hoc} parsers for the process of transforming XML files to RDF. 

In future work, we will further investigate the use of mapping languages in terms of its performance when considering large documents. Future work will also investigate the performance of the RDF knowledge graph based on queries, which may result in improvements on the semantic model, and in changes in the mapping.

\section*{Acknowledgments}
This paper was supported by the Science Foundation Ireland (Grant 13/RC/2106) as part of the ADAPT Centre for Digital Content Technology (\url{http://www.adaptcentre.ie/}) at Trinity College Dublin.

\bibliographystyle{splncs04}
\bibliography{ai4legal-paper}
\end{document}